# DESIGN AND FABRICATION OF LOW COST INTELLIGENT CARDIOPULMONARY RESUSCITAION DEVICE

Nabeel Ahmad Khan Jadoon[1], Shahzad Anwar[2]

[1][2]Department of Mechatronics Engineering,

University of Engineering and Technology, Peshawar

Author's Email: nabeeljadoon21@gmail.com

**ABSTRACT**: Cardiac arrest is a prevalent medical emergency that requires immediate intervention to restore blood circulation and prevent fatal outcomes. Sudden cardiac arrest occurs when the heart abruptly stops beating, leading to a critical reduction in blood flow to vital organs, particularly the brain. Cardiopulmonary Resuscitation (CPR) is the primary life-saving procedure in such emergencies. However, maintaining an optimal and consistent compression rate manually is challenging, necessitating the development of mechanical CPR devices. This study proposes the design and fabrication of a cost-effective, automated Cardiopulmonary Resuscitation (CPR) device based on the principles of both the cardiac pump and thoracic pump mechanisms. The cardiac pump mechanism generates blood flow by compressing the sternum, while the thoracic pump mechanism enhances intrathoracic pressure through the elastic recoil of the ribs. The integration of these two concepts ensures an efficient and reliable resuscitation process. To meet the American Heart Association (AHA) guidelines, a closed-loop control system with a real-time feedback mechanism is implemented, optimizing compression depth and rate. Additionally, a Human-Machine Interface (HMI) is incorporated, allowing the device to be adapted for different age groups, including children, adults, and senior citizens. This intelligent system enhances usability and precision, making it a versatile solution for emergency medical response. The proposed system ensures effective resuscitation with minimal human effort, significantly improving survival rates in cardiac arrest cases.

*Note: This research requires medical certification hence it is not clinically approved yet.*

## I. INTRODUCTION

In recent surveys of American Heart Association showed that more than 93% died each year due to cardiac arrest and hardly 2.3% out of them get the resuscitation [1]. Concerning this American Heart Association (AHA) proposed the standard CPR guidelines in 2010. These guidelines told that compression rate needed for the patient should be 100 compression/min and depth of compression must be [1, 1.5inches] applied at the rate of 30:2. It means after 30 compressions, resuscitator need to provide two ventilation to the cardiac arrest victim. Afterward, based on those guidelines various device has been proposed for resuscitation procedure. Cardiopulmonary Resuscitation is an emergency procedure that combines the chest compression to provide the blood circulation of human body. Based on that principal different research and developments have been proposed and still on work. While process of CPR includes Coronary perfusion pressure acts as a reliable measure of predict the return of spontaneous circulation (ROSC). Furthermore, CPP is the difference between aortic the diastolic pressure and right atrial pressure of heart a CPP of 15 mmHg is needed. The rise in diastolic blood pressure reflects an increase in coronary flow while the highness in mean blood pressure reflects a positive impact on neural function. Leroy in 1829, proposed the method of ventilation using manual chest compressions with movements of hands [2]. Although manual heart compression mechanism was stablished for resuscitation but soon, they realized that human could not be able to provide 100compression/min at standard 30:1 ratio [3]. It implied, because human got exhausted and required depth of compression was also main enigma.

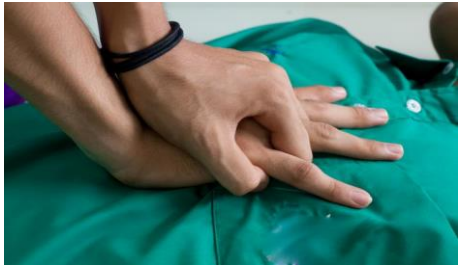

Fig 1: Chest compression by hand

In 1960 the paper by Jude and Knickerbocker described a method that required the little expertise [4]. Due to mechanical devices blood pressure approaches normal pressure. Unfortunately, standard closed-chest cardiopulmonary resuscitation (CPR) generates only 15-30% of the normal cardiac output. Coronary perfusion pressure above 20 mmHg is required to restore blood circulation [5]. Inconsistency of the current CPR technique persuade the investigators to obtain new CPR technique. In concern, there are two the mechanisms of for augmentation of blood flow. The cardiac pump and thoracic pump [6][7] enable the blood flow by compressing the heart and thorax between the sternums. Importantly, coronary perfusion pressure can be enhanced by cardiac pump. On the other hand thorax pump is able to increase the intrathoracic pressure. Hence, dual mechanism CPR device has been proposed which had tested and implemented which provides, 100compression/min standard rate of compression and flexible to implement on child's, Adults and Senior citizens.

## II. LITERATURE REVIEW

Regarding background of mechanical devices many CPR devices have been introduced since humans were unable to provide standard compression. Some of devices are as under on which this CPR device was based upon, The Thumper (Automatic Resuscitator), and the LUCAS device.

### A. The Thumper Device

The Thumper or Automatic Resuscitator is used mostly for CPR processes. Upper housing of this device is attached to board. It can provide 50% duty cycle rate [8]. The plunger is attached to the device, and it is pneumatic control [9]. The plunger's depth of compression is 1.5 to 2 inches can be adjusted by rescuer. The main disadvantage of Thumper was its costs [10]. Secondly, its pneumatic control mechanism was difficult to control.

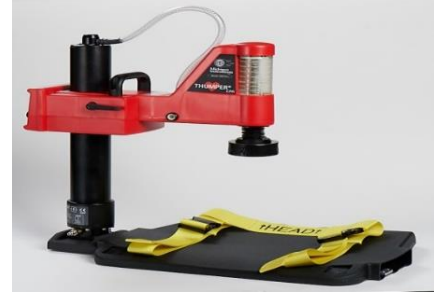

Fig 2: The Thumper Device

### B. LUCAS Device

The LUCAS Chest Compression System which was powered by battery [11]. This device was used to achieve 100 compression/min at 5cm depth of compression. The main disadvantages of this device are: It was not applicable to different age of patients. Its compression depth is fixed and more important design issue and operating issue as well. All that matter is sufficient ventilation ratio during resuscitation, but LUCAS device has no ventilation mask.

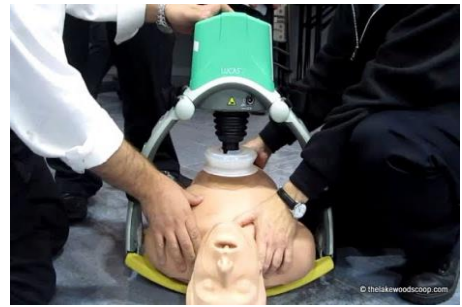

Fig 3: Lucas Device

## III. METHODOLOGY

### A. Simultaneous Sterno-Thoracic CPR: (SST-CPR)

Simultaneous Sterno thoracic CPR device was the dual induction mechanism in which heart act as cardiac pump and thorax cavity act as thoracic pump. Although single mechanism devices were there which constituted either of one mechanism but their survival rate was too low showed by the pilot study [12]. This mechanism was introduced to enhance the augmentation of blood flow and hemodynamic cycle of human body. In this device piston and constricting belts are used to achieve the dual mechanism process. In this regard, SST-CPR [12] increase the intrathoracic pressure with simultaneous compression of heart producing a rapid increase in blood flow. To more specify the process, firstly device CAD modelled was created in SolidWorks and simulated. Moreover, to justify the design of device, 3D printed the device. In order to make a CAD model for CPR device, there were main housing unit, piston assembly, supporting legs, and suction cup which were created as 3D model under require specifications and dimensions. Despite of all facts, its unique rolling mechanism was design to obtain the thoracic pump features. There were four rollers designed slightly offset from their sides regarding upward and downward position. While shown in CAD model, a belt was passed over the roller like a pulley; when piston did, compression and decompression restricting belts were moved along the vice versa motion of piston. In this way, dual mechanism of CPR device was achieved. With supporting legs, tappers were attached so that this device could be adjusted to any size of patient, right above the patient chest

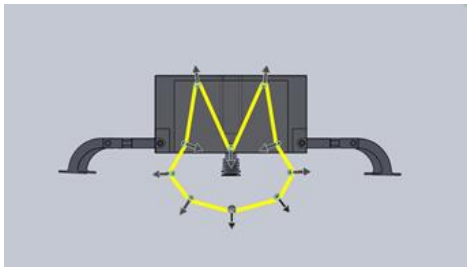

Fig 4: CAD Model

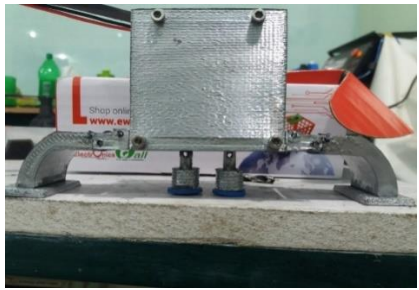

Fig 5: 3D printing of CPR Device

### B. Dimensions of Device

To obtain and calculate the required specified dimensions. Simulation on SolidWorks software has been done. Moreover, stress analysis and moment of Inertia for compressing mechanism was calculated and which matched with the standard specifications of American Heart Association. Regarding specifications, below table shows the result. This CPR device is based on mentioned specifications in Table 1 for its fabrication.

Table 1: Description of Size and Material of CPR Device

| COMPONENTS | DIMENSIONS L x W x H (inches) | NO | MATERIALS |
|---|---|---|---|
| PISTON CYLINDER | 13 x 0.5<br>10 x 2 x 2 | 2<br>2 | ALUMINIUM |
| FRAME | 16 x 5 x 8 | 1 | ALUMINIUM |
| BELT | 25 x 4 x 0.5 | 1 | POLYESTER |
| ROLLER | 5 x 0.5 | 4 | METAL ROD |
| SUPPORTING LEGS | 20 x 0.5 | 4 | IRON |

### IV. FABRICATION

Fabrication process needed much attention and specifications of material. Previously, stress analysis was performed as Aluminium (Al) was selected as prime part. To fabricate the compressing mechanism of piston, secondary material (Fe) Iron is used. Particularly, sliders, disc and frame but for piston rod TEFLON material had used [13]. Moreover, other components of the device are upper assembly, electronic housing, supporting legs, rollers, tappers and HMI feedback system.

#### A. Compressing Mechanism

Compressing mechanism was fabricated as Cam follower mechanism as shown in figure. To mechanize the system, a disc was attached to motor which rotates along the slider and rotatory motion was converted into linear motion. Although friction force (Fe) considered while fabricating the system. To minimize the friction force, lubricating fluid was used such as machine oil, Greece etc. It is shown in disc there are too many holes, these holes were pored for contingency procedure. For instance, if a device control system may fail, this device could be used to adjust the depth rate needed for standard compression.

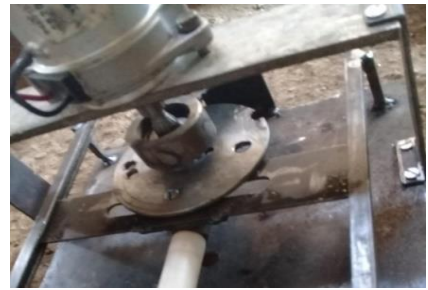

Fig 6: Compressing Mechanism of CPR

#### B. Electronics Housing

E-Housing of CPR device was specifically selected for electronics components and controller, power supplies and motor shield. In order to insulate all the process separate apartment was needed for the CPR device.

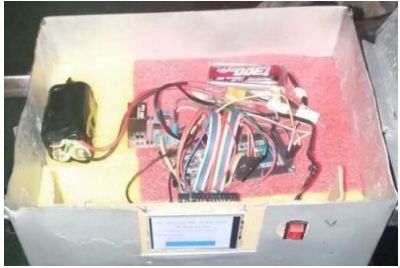

Fig 7: Electronics Housing

Fabrication of full assembly of device had completed and tested successfully to obtain the required results. To control the device a small HMI (Human machine interface) was developed select the type of patient, and to ensure the right compression rate and compression depth would be provided to the patient.

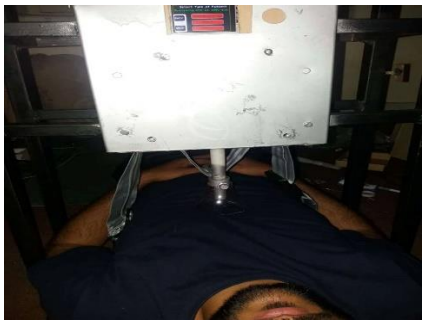

Fig 8: Full Assembly of CPR Device

## V. RESULTS

### A. Fuzzy Logic on CPR Device

Fuzzy logic implementation has been performed on device to obtain the effective and accurate results, which are in form of piston compression rate and compression depth control, using the age factor (of Adults, Child and Senior Citizens) as our input for the device.

### B. Fuzzy Logic Control

Fuzzy logic is soft computing technique used in control of machines [14]. The term "fuzzy" indicates the human approximation that can deal with concepts of artificial intelligence and soft computing [28]. Fuzzy logic has the significance that problem is more understandable by humans. Human experience and knowledge base can be used in the design of the controller. This makes it a lot easier to handle tasks that are performed by humans.

Input/output listing of device for fuzzy logic control were Age as primarily factor input, and Compression depth and rate as output variable.

### C. Inputs

Table 2: Fuzzy Logic Inputs Specification for Membership functions

| Patients | Fuzzy sets | Range of Functions |
|---|---|---|
| Senior citizen | Fuzzy set: SC= {x1, µsc(x1)} | Range: x1 € [38<x1<=70] , µsc € [0, 1] |
| Adult: | Fuzzy set: SC= {x2, µA(x1)} | Range: x2€ [18<x2<=40] , µA € [0, 1] |
| Child: | Fuzzy set: SC= {x3, µA(x1)} | Range: x3€ [10<x3<=20] , µc € [0, 1] |

Based on Table 2, input membership functions were defined for child, adult and senior citizen as shown in fig 9. Regarding fuzzy logic there must be intersection between the ranges, intersection was introduced so that membership grade possibility of occurring can be determined.

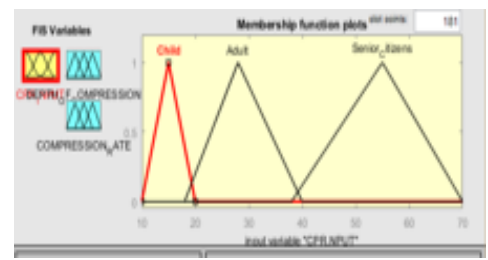

Fig 9: Input membership functions

### D. Outputs

Table 3: This Table shows the depth in inches and compression rate/min.

| Input Factor | Depth of Compression (in) | Compression Rate (/min) |
|---|---|---|
| Child | 0.8-0.95 | 70-85 |
| Adult | 1.20-1.50 | 105-120 |
| Senior Citizen | 0.92-1.15 | 80-110 |

Concerning output, triangular membership functions were also defined based on Table 3 data. Moreover, compression rate and compression depth membership functions for three types of patients were so that possibility of occurring one of these values could be easily obtained and use for CPR device.

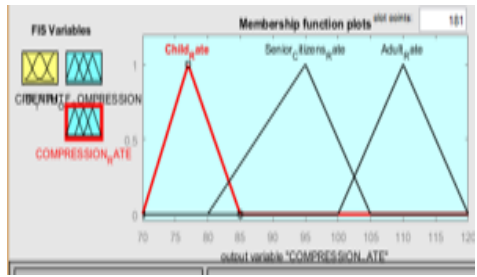

Fig 10: Compression rate membership function

Membership functions for depth of compression were also defined in fuzzy logic MATLAB Toolbox as shown in fig 11.

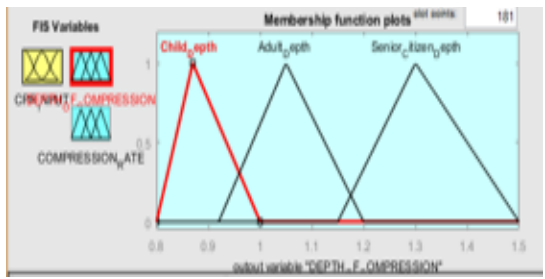

Fig 11: Depth of compression Membership function

### a. FUZZIFICATION PROCESS

In order to fuzzified the problem, we have used the Mamdani rule of composition [15]. Which select the MIN and MAX combination of rules which means, intersection and union respectively. Moreover, for rule of composition another term also used, which is Inference Engine. This engine matches the required rule to be executed first and fired the that rule to obtain the result [16][17].

Regarding fuzzy process for CPR machine, three rules have been defined for three types of patients. Importantly, there must be intersection between the inputs so that human approximation and soft computing method can be implemented. These values are linguistic. Sometimes, these are called as fuzzy descriptors. In order to convert it into real values, defuzzification process was used to obtain the required results.

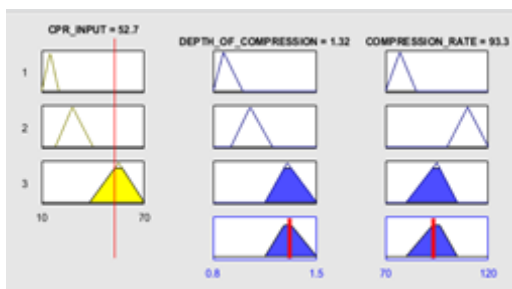

Fig 12: Fuzzification process on MATLAB

### b. Defuzzification process

This was the most important phase to obtain the real result in form of crisp values [18]. Crisps are those value which are either "true" or "false". We have used the centroid method to defuzzify the output membership number for obtaining the pre-requisites. Particularly, MATLAB result shown, was same as the actual one. For defuzzification compression rate and CPR input were kept along y-axis and x-axis respectively. As the range for different age of patients was selected from [10, 70]. If the input was primary the age factor chosen between this range, then the compression rate would be provided to victims of cardiac arrest.

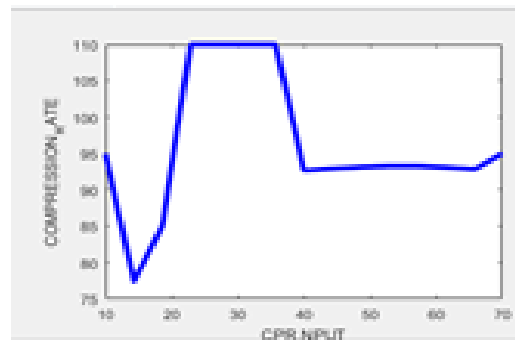

Fig 13: Compression Rate

Secondly, this graph indicates the depth of compression neeeded for pateients. Input was same for depth of compression but range of depth required for adult, child and senior citizens was defined.

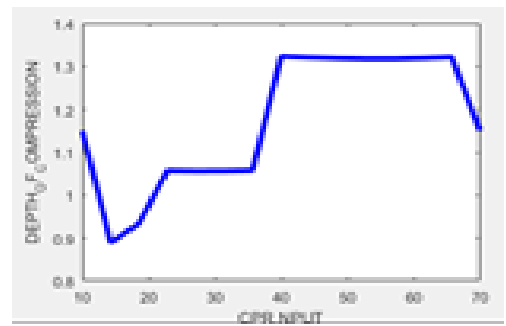

Fig 14: Depth of Compression

## VI. CONCLUSION AND FUTURE RESEARCH

Cardiopulmonary Resuscitation (CPR) is a critical emergency procedure that has been successfully implemented in this device using a dual-mechanism theory. This mechanism enhances the efficiency of chest compressions by integrating both cardiac pump and thoracic pump principles, ensuring improved blood flow during resuscitation. The device's compression rate and depth are precisely controlled through an integrated feedback system, allowing real-time adjustments for different patient categories. A small Human-Machine Interface (HMI) provides real-time input parameters, making the device adaptable for children, adults, and senior citizens.

To ensure optimal performance, the compression rate and depth values are pre-calculated and programmed into the device, enabling accurate and effective resuscitation. Compared to existing devices such as LUCAS CPR and Zoll CPR, this system offers enhanced blood flow augmentation due to its unique dual-mechanism approach. The device has significant applications in hospitals, ambulances, and first-aid services, where rapid and effective CPR is crucial.

During the fabrication process, two primary challenges were encountered:

1. Device weight – Efforts were made to minimize the weight, but further optimization is required to improve portability.

2. Motor-induced vibrations – The current compression mechanism generates vibrations due to friction with sliders. To address this, an alternative mechanism could be explored to eliminate vibrations, enhancing overall stability and efficiency.

By overcoming these challenges, the device can be further refined to ensure greater reliability, portability, and effectiveness in emergency medical scenarios.

## VII.   ACKNOWLEGMENT

We pay our gratitude to Directorate of Science and Technology (DoST) foundation Pakistan for funding this project.

## REFERENCES

[1] M. W. Cooke and G. Perkin, "Variability in cardiac arrest survival: the NHS Ambulance Service Quality Indicators," *Emerg. Med. J.*, vol. 29, pp. 3–5, 2012. [2]. Brooks. Hassan, N., Bigham, B.L., Morrison, L.J. Mechanical versus manual chest compressions for cardiac arrest. Cochrane Database System Rev. 2014; 2: CD007260.

[2] N. Hassan Brooks, B. L. Bigham, and L. J. Morrison, "Mechanical versus manual chest compressions for cardiac arrest," *Cochrane Database Syst. Rev.*, vol. 2, Art. no. CD007260, 20[3] "Guideline for cardiopulmonary resuscitation: Recommendation of the 1992 National Conference, Emergency Cardiac Care Committee and Subcommittees, American Heart Association," *JAMA*, vol. 268, pp. 2171–2302, 1992.

[4] K. Knickerbocker, W. B. Kouwenhoven, and J. R. Jude, "Closed-chest cardiac massage," *JAMA*, vol. 173, pp. 1064–1067, 1960.

[5] W. B. Kouwenhoven, J. R. Jude, and G. G. Knickerbocker, "Closed chest cardiac massage," *JAMA*, vol. 173, pp. 1064, 1960.

[6] J. A. Warner, H. L. Green, C. L. Janko, et al., "Visualization of cardiac valve motion in man during external chest compression: Implications regarding the mechanism of blood flow," *Circulation*, vol. 63, pp. 1417–1421, 1981.

[7] M. Antz and M. Matthias, "Augmentation of blood flow circulation and techniques," *Circulation*, vol. 98, no. 17, pp. 1790–1795, 1998.

[8] A. Norman, S. Lemeshow, and J. Brewer, "Inhomogeneity and temporal effects in Auto Pulse Assisted International Resuscitation—an exception from experimental trial terminated early," *Am. J. Emerg. Med.*, vol. 28, no. 4, pp. 391–398, 2010.

[9] J. T. Nieman, J. P. Roseboro, M. Bauknecht, et al., "Pressure-synchronized cineangiography during experimental cardiopulmonary resuscitation," *Circulation*, vol. 64, pp. 985–991, 1981.

[10] M. P. Feneley, G. W. Maier, K. B. Kem, et al., "Influence of compression rate on initial success of resuscitation and 24-hour survival after prolonged manual cardiopulmonary resuscitation in dogs," *Circulation*, vol. 77, pp. 240–250, 1988.

[11] T. P. Aufderheide, R. J. Francesco Cone, M. A. Wayne, et al., "Cardiopulmonary resuscitation versus active compression-decompression cardiopulmonary resuscitation: A randomized trial," *Lancet*, vol. 377, pp. 301–311, 2011.

[12] M. T. Rudik, W. L. Maughamium, M. Ephron, P. Freund, and M. L. Westfeldt, "Mechanisms of blood flow during cardiopulmonary resuscitation," *Circulation*, vol. 61, pp. 345–352, 1980.

[13] J. M. Plunket, "Tetrafluoroethylene polymers," U.S. Patent 4, issued Feb. 4, 1941.

[14] S. K. Bezdek and S. Plate, *Fuzzy Models for Pattern Recognition*, New York, NY, USA: IEEE Press, 1992.

[15] M. Drobics and U. Bodenfire, "Fuzzy logic modeling with human approximation," in *Proc. 2002 IEEE Int. Conf. Syst., Man, Cybern.*, vol. 4, Hammamet, Tunisia, 2002.

[16] D. Nakul, Frankenstein, and R. Kruse, *Fuzzy-System and Computational Intelligence*, 2nd ed., vol. 2–3, Braunschweig, Germany: Vieweg, 1996.

[17] L. A. Zadeh and Mamdani, "A new approach to the analysis and optimization of complex systems

and decision processes," *IEEE Trans. Syst., Man, Cybern.*, vol. 3, no. 1, pp. 28–44, 1973.

**[18]** J. W. Coupland, "A centroid method for defuzzification of type-2 fuzzy sets," *IEEE Trans. Fuzzy Syst.*, vol. 16, no. 4, pp. 929–941, 2008.